\documentclass[12pt,a4paper]{article}
\pdfoutput=1

\usepackage{epsfig}
\usepackage{amsmath}
\usepackage{amssymb}
\usepackage{verbatim}
\usepackage{bm}
\usepackage{dsfont}
\usepackage[footnotesize]{caption}
\usepackage{subfigure}
\usepackage{cancel}
\usepackage{cite}
\usepackage{a4wide}
\usepackage{color}
\usepackage{hyperref}
\usepackage{multirow}
\usepackage{url}
\usepackage[utf8]{inputenc}

\hypersetup{
    colorlinks = true,
    linkcolor = blue,
    linkbordercolor = {1 0 0}, % RGB values for red (1 0 0)
    citecolor = blue,
    urlcolor = blue
}

\usepackage{geometry}
\renewcommand{\baselinestretch}{1.2}

\begin{document}

\begin{minipage}{0.4\textwidth}
\begin{flushleft}
CERN-TH-2023-118 \\
MS-TP-23-37 \\
DESY-23-117
\end{flushleft}
\end{minipage} \hfill
\begin{minipage}{0.4\textwidth}
\begin{flushright}
 July 2023 \\
 \phantom{xx} \\
 \phantom{xx}
\end{flushright}
 \end{minipage}

\vskip 2.5cm

\begin{center}

{\Large\bf Metastable cosmic strings}

\vskip 1cm

{\large Wilfried~Buchm\"uller$^a$, Valerie~Domcke$^b$,  Kai~Schmitz$^c$
}\\[3mm]
{\it 
$^a$ Deutsches Elektronen-Synchrotron DESY, 22607 Hamburg, Germany\\
$^b$ Theoretical Physics Department, CERN, 1211 Geneva 23, Switzerland \\
$^c$ Institute for Theoretical Physics, University of M\"unster, 48149 M\"unster, Germany
}
\end{center}

\vskip 2cm

\begin{abstract}
\noindent 
Many symmetry breaking patterns in grand unified theories (GUTs) give rise to cosmic strings that eventually decay when pairs of GUT monopoles spontaneously nucleate along the string cores. These strings are known as metastable cosmic strings and have intriguing implications for particle physics and cosmology. In this article, we discuss the current status of metastable cosmic strings, with a focus on possible GUT embeddings and connections to inflation, neutrinos, and gravitational waves (GWs). The GW signal emitted by a network of metastable cosmic strings in the early universe differs, in particular, from the signal emitted by topologically stable strings by a suppression at low frequencies. Therefore, if the underlying symmetry breaking scale is close to the GUT scale, the resulting GW spectrum can be accessible at current ground-based interferometers as well as at future space-based interferometers, such as LISA, and at the same time account for the signal in the most recent pulsar timing data sets. Metastable cosmic strings thus nourish the hope that future GW observations might shed light on fundamental physics close to the GUT scale.
\end{abstract}

\thispagestyle{empty}

\newpage
{\hypersetup{linkcolor=black}\renewcommand{\baselinestretch}{1}\tableofcontents}

\section{Introduction}
\label{sec:intro}

The formation of topological defects is a generic feature of cosmological phase transitions~\cite{Kibble:1976sj}.
Such defects are tied to spontaneous symmetry breaking in extensions of the Standard Model (SM), in particular in grand
unified theories (GUTs). They include Nielsen--Olesen strings~\cite{Nielsen:1973cs}, 't Hooft--Polyakov monopoles~\cite{tHooft:1974kcl,Polyakov:1974ek}, unstable ``dumbbells'' or ``X-strings'' connecting a monopole--antimonopole pair~\cite{Nambu:1977ag}, and other composite defects~\cite{Kibble:1982ae}.

Monopoles would overclose the universe and must therefore be avoided or diluted by inflation. Domain walls will reach a scaling regime, but will still lead to an overclosure problem. On the contrary, cosmic strings evolve towards a scaling regime where their fraction of the total energy density remains constant. Together with characteristic signatures in the cosmic microwave background and in gravitational lensing, the stochastic gravitational-wave background (SGWB) from cosmic strings is a potentially very interesting messenger from the early universe (for reviews and references, see, e.g., Refs.~\cite{Vilenkin:2000jqa,Hindmarsh:2011qj,Auclair:2019wcv}).

For a large class of supersymmetric GUTs with symmetry breaking chains avoiding the monopole problem, cosmic-string formation is unavoidable~\cite{Jeannerot:2003qv}. Making use of supersymmetric hybrid inflation~\cite{Copeland:1994vg,Dvali:1994ms}, the string scale is close to the GUT scale, a prominent example being the breaking of $B\!-\!L$, the difference between baryon and lepton number~\cite{Buchmuller:2012wn}. String scales below the GUT scale are also possible and may be related to intermediate-mass right-handed neutrinos, which could render the SGWB a probe of thermal leptogenesis~\cite{Dror:2019syi}.

Pulsar timing array (PTA) observations~\cite{Arzoumanian:2018saf,Kerr:2020qdo,Shannon:2015ect} can probe the string tension of stable cosmic strings down to $G\mu \lesssim 10^{-10}$~\cite{Blanco-Pillado:2017rnf}, where $G$ denotes Newton's constant and $\mu$ is the energy per unit length of the string. These observations have now entered a new phase with evidence for a common-spectrum process at nanohertz frequencies first reported in Refs.~\cite{Arzoumanian:2020vkk,Goncharov:2021oub,Chen:2021rqp,Antoniadis:2022pcn}, followed by evidence for  Hellings--Downs angular correlation, the smoking-gun signal of a SGWB, reported by PTA collaborations across the world in Refs.~\cite{NANOGrav:2023gor,Antoniadis:2023ott,Reardon:2023gzh,Xu:2023wog}. 
Beyond the astrophysical interpretation in terms of inspiraling supermassive black-hole binaries~\cite{Middleton:2020asl,NANOGrav:2023hfp,Antoniadis:2023xlr}, possible cosmological interpretations include stable and metastable cosmic strings~\cite{Ellis:2020ena,Blasi:2020mfx,Buchmuller:2020lbh,Bian:2020urb,Blanco-Pillado:2021ygr,EPTA:2023hof} (see, e.g., Refs.~\cite{Madge:2023cak,NANOGrav:2023hvm,Antoniadis:2023xlr} for an overview of possible cosmological signals). However, the originally favoured GUT-scale strings with a tension in the range $G\mu \simeq 10^{-(8\cdots6)}$ are firmly excluded by these results, as they would lead to too large an SGWB signal in the PTA band.

However, in theories where strings couple to monopoles, strings can decay by quantum tunneling into string segments connecting monopole--antimonopole pairs~\cite{Vilenkin:1982hm}. In the semiclassical approximation, the decay rate per string unit length is given by~\cite{Preskill:1992ck,Leblond:2009fq,Monin:2008mp}
\begin{equation}\label{decayrate}
\Gamma_d = \frac{\mu}{2 \pi} \exp\left( - \pi \kappa \right) \qquad \text{with} \qquad 
\kappa = \frac{m_M^2}{\mu}  \,,
\end{equation}
where $m_M$  is the monopole mass. Given the exponential dependence of the decay rate on the parameter $\kappa$, and considering monopole masses larger than the string scale, metastable strings have generally been assumed to be effectively stable (see, e.g., Refs.~\cite{Vilenkin:1982hm,Vilenkin:2000jqa,Dror:2019syi}). However, a particularly interesting phenomenology is obtained for metastable cosmic strings with $\sqrt{\kappa} \sim 8$. Such values can indeed be obtained for $\text{SO}(10)$ models with $B\!-\!L$ strings~\cite{Buchmuller:2021dtt}. In this case, the cosmic-string network survives for about an hour (redshift $z \sim 10^7$) until monopole production becomes efficient, which implies that at high frequencies the resulting GW spectrum resembles that of stable cosmic strings, whereas at lower frequencies, corresponding to GWs sourced at later times, the spectrum is strongly suppressed~\cite{Buchmuller:2019gfy,Gouttenoire:2019kij,Dunsky:2021tih}. As a result, metastable cosmic strings can not only provide a good fit to the PTA signal for GUT-scale string tensions~\cite{Buchmuller:2020lbh,NANOGrav:2023hvm}, but moreover, they can also easily evade any bounds at PTA scales while still yielding a strong signal at higher frequencies~\cite{Buchmuller:2019gfy}, i.e., in the frequency bands relevant for LISA~\cite{Audley:2017drz} and ground-based interferometers~\cite{LIGOScientific:2014pky, VIRGO:2014yos, KAGRA:2020tym}.

In this article, we will discuss the current status of metastable cosmic strings. Section~\ref{sec:metastrings} deals with a minimal but representative example model: the breaking of $\text{SU}(2)_R\times\text{U}(1)_{B-L}$ down to $\text{U}(1)_Y$ by an $\text{SU}(2)_R$ Higgs triplet and two $\text{SU}(2)_R$ Higgs doublets with quantum numbers suitable for an embedding in $\text{SO}(10)$. The computation of the SGWB signal is presented in Section~\ref{sec:sgwb}, with an emphasis on the theoretical prediction for the spectral tilt of the GW spectrum. Some aspects of stable and quasi-stable strings are reviewed in Section~\ref{sec:qscs}, and the role of inflation is described in Section~\ref{sec:inflation}. We conclude in Section~\ref{sec:conclusions}.

\section{Metastable strings}
\label{sec:metastrings}

Metastable strings are a characteristic prediction of GUTs that lead, via several steps of spontaneous symmetry breaking, to the SM gauge group $\text{G}_{SM} = \text{SU}(3)_C\times\text{SU}(2)_{L}\times\text{U}(1)_{Y}$. Strings with tensions above the electroweak scale result from the spontaneous breaking of a $\text{U}(1)$ group that commutes with $\text{G}_{SM}$. Similarly, monopoles arise once a non-Abelian gauge group is broken to a subgroup containing a $\text{U}(1)$ factor. Then, if the $\text{U}(1)$ symmetry involved in the production of monopoles partially overlaps or coincides with the $\text{U}(1)$ symmetry responsible for string formation, the strings become metastable, i.e., pairs of monopoles and antimonopoles spontaneously nucleate along the strings by quantum tunneling.

SM extensions giving rise to strings must feature a gauge group of at least rank $5$. Starting from an exceptional Lie group at high energies, we can, e.g., consider the following symmetry breaking chain,
\begin{equation}\label{embSU5}
  \text{G}_{SM}
  \subset \text{SU}(5)\times\text{U}(1)_{X} 
  \subset \text{SO}(10) \subset \text{E}(6) \subset \dots \,,
\end{equation}
where $\text{SU}(5)$ refers to the Georgi--Glashow
$\text{SU}(5)$ GUT group or to the flipped $\text{SU}(5)$
model. Another possibility is to consider a sequence
featuring an extended electroweak sector, 
\begin{equation}\label{embLR}
  \begin{split}
    \text{G}_{SM}
    %&= \text{SU}(3)_C\times\text{SU}(2)_{L}\times\text{U}(1)_{Y}
  \subset \text{SU}(3)_C\times\text{SU}(2)_{L}\times\text{SU}(2)_{R}
  \times\text{U}(1)_{B-L} 
   \subset
  \text{G}_{PS}\subset \text{SO}(10) \subset \text{E}(6) \subset \dots \,,
\end{split}
\end{equation}
where $\text{G}_{PS} = \text{SU}(4)\times\text{SU}(2)_{L}
\times\text{SU}(2)_{R}$ denotes the Pati--Salam group.
  
If a symmetry group $\text{G}$  is broken to a subgroup
$\text{H}$, the quotient $\mathcal{M} = \text{G}/\text{H}$
corresponds to the manifold of degenerate vacuum states.
The types of defects that may be formed in the symmetry breaking are
governed by the topology of $\mathcal{M}$, which is encoded in the
homotopy groups $\pi_n(\mathcal{M})$. Topologically stable
strings can form if the first homotopy group is nontrivial,
$\pi_1(\mathcal{M})  \neq I$, i.e., there are loops in $\mathcal{M}$ that cannot be
contracted to a point. Similarly, topologically stable magnetic monopoles can arise if the
second homotopy group is nontrivial, $\pi_2(\mathcal{M}) \neq I$, so
that there exist non-contractable two-dimensional surfaces in $\mathcal{M}$.
We shall be particularly interested in two-step symmetry breakings
$\text{G} \rightarrow \text{H} \rightarrow \text{K}$, where the
homotopy group $\text{G}/\text{K}$ is trivial, but the homotopy groups
of the individual steps, $\text{G}/\text{H}$ and
$\text{H}/\text{K}$, are nontrivial. In this case, metastable defects can form.

A simple example is the breaking of $\text{SO}(10)$ to the Standard
Model group via $\text{SU}(5)$. The result crucially depends on the
chosen Higgs representation \cite{Kibble:1982ae}. The breaking chain
\begin{equation}\label{SU5A}
  \text{SO}(10) \stackrel{\bf{45}}{\rightarrow} \text{SU}(5) \times
  \text{U}(1) \stackrel{\bf{45}\oplus\bf{126}}{\rightarrow}
  \text{G}_{SM} \times \mathbb{Z}_2
\end{equation}
yields stable monopoles and, in the second step, also stable strings.
On the contrary, for the closely related symmetry
breaking with a $\bf{16}$-plet,
\begin{equation}\label{SU5B}
  \text{SO}(10) \stackrel{\bf{45}}{\rightarrow} \text{SU}(5) \times
  \text{U}(1) \stackrel{\bf{45}\oplus\bf{16}}{\rightarrow}
  \text{G}_{SM} \,,
\end{equation}
the homotopy group of $\mathcal{M} = \text{SO}(10)/\text{G}_{SM}$ is trivial, $\pi_1(\mathcal{M})  = I$, and there
are no topologically stable strings. However, 
cosmologically interesting metastable strings can now form.

Metastable strings can break apart into segments in consequence of
quantum tunneling events leading to the spontaneous nucleation of
monopole--antimonopole pairs. Eventually, string decay leads to a
population of short string segments where each segment has a monopole
on one end and an antimonopole on the other. In the example in Eq.~\eqref{SU5B},
monopoles are formed both in the first and second breaking step,
where the latter also determines the string energy scale.
Besides, there are also other composite topological defects, which can be created in other symmetry breaking chains. One example are $\mathbb{Z}_2$-strings, also known as ``necklaces'', which correspond to one-dimensional string--monopole--string configurations, i.e., configurations where two strings are attached to each monopole~\cite{Hindmarsh:1985xc,Leblond:2007tf}. More details and references on composite topological defects can be found in Ref.~\cite{Kibble:2015twa,Dunsky:2021tih,Lazarides:2023iim}.

Realistic GUTs require large Higgs representations in order to break the GUT gauge group down to the SM, which complicates their analysis. On top, nonsupersymmetric models are sensitive to large radiative corrections and hence suffer from a severe naturalness problem. This observation triggered the investigation of even more complicated models in the literature: supersymmetric GUTs with even more complicated Higgs sectors. In view of this situation, it is important not to loose sight of the fact that conventional spontaneous symmetry breaking is not the only way in which a fundamental GUT gauge group can be reduced to the SM gauge group. Higher-dimensional theories such as orbifold GUTs or string theory represent intriguing alternatives that deserve consideration (a review and references can, e.g., be found in Ref.~\cite{Raby:2017ucc}). In these constructions, the fundamental GUT gauge group is first partially broken in a geometric way, namely, by the compactification of extra dimensions, and only the remnant subgroup remaining after this first step is further reduced to the SM group via conventional spontaneous symmetry breaking. For these reasons, we will restrict ourselves to the simplest possible case leading to metastable strings in the following, the first embedding in Eq.~\eqref{embLR}, which may mark the end of a long symmetry breaking chain, which we, however, do not specify in detail,
\begin{equation}\label{embU2}
    \text{G}_{SM} \subset \text{SU}(3)_C\times\text{SU}(2)_{L}\times\text{U}(2) \ ,
  \quad \text{U}(2) = \text{SU}(2)_{R}\times\text{U}(1)_{B-L} /\mathbb{Z}_2 \ .
\end{equation}
In order to break this group down to $\text{G}_{SM}$, we consider a Higgs triplet $U \sim (3,0)$ of $\text{SU}(2)_{R}$ alongside a pair of Higgs doublets of $\text{SU}(2)_{R}$, $S \sim (2,q)$ and $S_c \sim (\bar{2},-q)$, that carry charges $\pm q$ under $\text{U}(1)_{B-L}$. 
The breaking of $\text{SU}(2)_{R}$ leads to monopoles while the breaking of $\text{U}(1)_{B-L}$ implies strings, yielding the necessary ingredients for metastable strings.
Also, note that we divide out a $\mathbb{Z}_2$ factor in Eq.~\eqref{embU2}, which is necessary to avoid double counting of the center of $\text{SU}(2)_{R}$, which consists of the identity element and its negative, $\left\{I,-I\right\}$, and which is also contained in $\text{U}(1)_{B-L}$. For earlier discussions of defects in $\text{U}(2)$ models with triplet and doublet Higgs fields but without supersymmetry, see Refs.~\cite{Copeland:1987ht,Kephart:1995cg,Achucarro:1999it,Kibble:2015twa}.

\subsection[Strings from supersymmetric $B\!-\!L$ breaking]{Strings from supersymmetric \boldmath{$B\!-\!L$} breaking}
\label{subsec:strings}

The prospects to explain the recent PTA signal in terms of a SGWB from metastable cosmic strings motivate us to consider large string tensions. We are specifically interested in symmetry breaking scale far above the electroweak scale, at least of the order of $v_s \sim 10^{13}~\text{GeV}$. It is reasonable to expect unbroken supersymmetry at such high energies, which is why we will focus on supersymmetric models of symmetry breaking in the rest of this paper, following the analysis presented in Ref.~\cite{Buchmuller:2021dtt}.

Our starting point is a supersymmetric Abelian Higgs model with two chiral superfields $S$ and $S_c$ and a gauge singlet $\phi$ that gives rise to spontaneous $B-L$ breaking. The fields $S$ and $S_c$ carry charge $q$ and $-q$ under $U(1)_{B-L}$, respectively, and the K\"ahler potential and superpotential of the model (we use the same conventions as in Ref.~\cite{Wess:1992cp}) are given by
\begin{equation}\label{KPU1}
K = S^\dagger e^{2gqV} S + S_c^\dagger e^{-2gqV}S_c + \phi^\dagger\phi \,,\qquad
P = \frac{1}{4}\,W W + \lambda\phi\left(\frac{v_s^2}{2} - SS_c\right) \,.
\end{equation}
Here, $V$ is a vector superfield, $W$ is the supersymmetric field strength, and $v_s$ is the scale of spontaneous symmetry breaking, which we can choose to be real and positive.
From the auxiliary fields of the vector and chiral superfields, we can derive the scalar potential,
\begin{equation}
\mathcal{V} = \frac{1}{2} D^2 + |F_S|^2 + |F_{S_c}|^2 + |F_\phi|^2 \ ,
\end{equation}
where the $F$ and $D$ terms follow from solving the associated equations of motion,
\begin{align}
D & = - gq \left(\left|S\right|^2 - \left|S_c\right|^2\right) \,, \\ 
F_S^* & = \lambda\phi S_c \,, \qquad  
F_{S_c}^* = \lambda\phi S \,, \qquad
F_\phi^* = -\lambda\left(\frac{v_s^2}{2} - SS_c\right) \,.
\end{align}
The scalar potential and the kinetic terms for the scalar and vector fields constitute the bosonic part of the Lagrangian,
\begin{equation}
\mathcal{L}_b = -\frac{1}{4} F_{\mu\nu}F^{\mu\nu} - (D_\mu S)^*(D^\mu S)
- (D_\mu S_c)^*(D^\mu S_c) - \partial_\mu \phi^* \partial^\mu \phi -
\mathcal{V} \ ,
\end{equation}
with covariant derivatives $D_\mu  = \left(\partial_\mu + igq A_\mu\right)S$ and $D_\mu S_c = \left(\partial_\mu - igq A_\mu\right)S_c$, and where $A_\mu$ is the vector component in the vector multiplet $V$, $F_{\mu\nu}$ is the corresponding field strength, and where chiral superfields and their scalar components are denoted by the same symbols.

The vacuum manifold of the model is described by a $D$-flat direction, $\left|S\right|^2 = \left|S_c\right|^2$, which represents a flat moduli space of vacuum states with unbroken supersymmetry and spontaneously broken $\text{U}(1)_{B-L}$ symmetry,
\begin{equation}
  S = \frac{v_s}{\sqrt{2}}\,e^{i\alpha}\ , \quad
  S_c = S^*\ , \quad \phi = 0\ .
\end{equation}
The particle excitations around the true vacuum are best described if we expand the fields $S$ and $S_c$ around their vacuum expectation values (VEVs),
\begin{equation}
S = \frac{v_s}{\sqrt{2}}\,e^{i\alpha} + S' \,, \qquad S_c = \frac{v_s}{\sqrt{2}}\,e^{-i\alpha} + S'_c \,.
\end{equation}
In this field basis, we then find that the Goldstone multiplet $\left(S'-S'_c\right)/\sqrt{2}$ is ``eaten'' by the massless vector multiplet $V$, which results in a massive vector multiplet with mass $m_V = \sqrt{2}gqv_s$. Similarly, the orthogonal linear combination $\left(S'+S'_c\right)/\sqrt{2}$ and the singlet field $\phi$ fuse in a massive chiral multiplet with mass $m_S = \lambda v_s$.

As in the nonsupersymmetric case, the vacuum manifold $\mathcal{M}$ is the circle $S^1$, which has a nontrivial first homotopy group, $\pi_1(\mathcal{M}) = \mathbb{Z}$. The model thus admits exited states in the form of topologically stable strings. On the supersymmetric moduli space, i.e., along the $D$-flat direction, these strings are described by the Nielsen--Olesen string solutions~\cite{Nielsen:1973cs}. Static strings along the $z$ axis and with winding number $n$ correspond to field configurations of the form
\begin{equation}
S = \frac{v_s}{\sqrt{2}} \, f\left(\rho\right) e^{ni\varphi} = S_c^* \,, \qquad
A_0 = 0 \,, \qquad A_i = -\frac{n}{g\rho} \, h\left(\rho\right) \partial_i\varphi \,,
\end{equation}
where we work in cylindrical coordinates $\left(\rho,\varphi,z\right)$ and with boundary conditions
\begin{equation}
f\left(0\right) = h\left(0\right) = 0 \,, \qquad f\left(\infty\right) = h\left(\infty\right) = 1 \,.
\end{equation}

Strings described by these field configurations exhibit a total magnetic flux along the string of $2n\,\pi/g$ with $n \in \mathbb{Z}\setminus\left\{0\right\}$ and a string tension (i.e., energy per unit length) of
\begin{equation}
\mu = 2\pi v_s^2\,B\left(\beta\right) = \frac{\pi m_V^2}{\left(gq\right)^2}\,B\left(\beta\right) \qquad\textrm{with}\qquad \beta = \frac{m_S^2}{m_V^2} = \frac{\lambda^2}{2(gq)^2} \,.
\end{equation}
Here, the parameter $\beta$ measures the ratio of the Higgs and vector boson masses; and $B$ is a slowly varying function of $\beta$, normalized such that $B\rightarrow1$ in the Bogomol'nyi limit $\beta \rightarrow 1$~\cite{Hindmarsh:2011qj}. If $\beta < 1$, the strings will have larger Higgs-field cores than gauge-field cores, $m_S^{-1} > m_V^{-1}$, which gives rise to type-I strings, in analogy to similar condensed-matter systems. Type-I strings are stable and attracted towards each other.

The model defined in Eq.~\eqref{KPU1} is intriguing, as it features a singlet field $\phi$ that can be identified as the inflaton in supersymmetric hybrid inflation \cite{Copeland:1994vg,Dvali:1994ms}. It is, moreover, straightforward to extend it by adding supersymmetry breaking terms and by coupling it to a set of chiral right-handed-neutrino superfields. Extensions of the model along these lines can account for leptogenesis, dark matter, and a stage of cosmic inflation in accord with recent bounds on the primordial scalar spectral index in observations of the cosmic microwave background~\cite{Buchmuller:2012wn,Buchmuller:2014epa}. Scenarios of this type also give rise to stable cosmic strings with tension $G\mu \sim 10^{-7}$, where the specific value of the string tension is dictated by the other phenomenological aspects of the model, in particular the energy scale of inflation. For stable strings, such large string tensions have, however, been known to be in conflict with PTA measurements for many years~\cite{Arzoumanian:2018saf,Kerr:2020qdo}, which renders such scenarios unviable.

\subsection{Monopoles from supersymmetric \boldmath{SU(2)} breaking}
\label{subsec:monopoles}

Next, we turn to topologically stable 't~Hooft--Polyakov monopoles \cite{tHooft:1974kcl,Polyakov:1974ek} produced by the spontaneous breaking of $\text{SU}(2)$ to $\text{U}(1)$. To implement this breaking, we consider an $\text{SU}(2)$ triplet $U^a$ ($a = 1,..,3$) and work with the following K\"ahler potential and superpotential,
\begin{equation}\label{KPSU2}
K = U^\dagger e^{2gV} U \,, \qquad
P = \frac{1}{8}\, \text{tr}\left[W W\right] \,.
\end{equation}
Here, $V = V^a T^a$ denotes the $\text{SU}(2)$ vector superfield and the $(T^a)_{bc} = -i\epsilon_{abc}$ are the $\text{SU}(2)$  generators in the adjoint representation. The bosonic Lagrangian of the theory is given by
\begin{equation}
\label{eq:Lagbos}
\mathcal{L}_b = -\frac{1}{4} F^a_{\mu\nu}F^{a\mu\nu} - (D_\mu U^a)^*(D^\mu U^a)
- ig \epsilon_{abc} D^a U^{b *} U^c + \frac{1}{2} D^aD^a +
F^{a *}_U F^a_U \,,
\end{equation}
with gauge-covariant derivative $\left(D_\mu U\right)^a = \partial_\mu U^a - g\,\epsilon_{abc}\, A^b_\mu U^c$ and non-Abelian field strength tensor $F^a_{\mu\nu} = \partial_\mu A^a_\nu - \partial_\nu A^a_\mu -g\,\epsilon_{abc}\,A^b_\mu A^c_{\nu}$.

In passing, we mention that the Lagrangian in Eq.~\eqref{eq:Lagbos} corresponds to the bosonic Lagrangian of $\text{SU}(2)$ Super-Yang--Mills theory with $\mathcal{N}=2$ supersymmetry. Supersymmetric UV completions of the SM of this type can occur in certain orbifold compactifications. Starting with a supersymmetric Pati--Salam or $\text{SO}(10)$ theory in five or six dimensions, orbifold constructions can lead to an $\mathcal{N} = 2$ sector with gauge group $\text{SU}(2)_R$ in four dimensions. The total gauge group containing $\text{SU}(2)_R$ as a subgroup is then spontaneously broken down to the SM in subsequent symmetry breaking steps.

The classical theory defined by the potentials in Eq.~\eqref{KPSU2} features again a moduli space spanned by the flat direction $U^a = u/\sqrt{2}\,\delta_{a3}$ modulo gauge transformations. Thanks to the large symmetry of the theory, this flat direction is preserved at the quantum level as well as when nonperturbative corrections are taken into account. The flat direction is therefore also present in the full theory, where it interpolates between two phases: a confinement phase with monopole condensation at small values of $u$ and a perturbative Higgs phase at large values of $u$~\cite{Seiberg:1994rs}. In the following, we will be concerned with the Higgs phase of the model, which corresponds to field values $u$ much larger than the confinement scale $\Lambda$.

As in the previous case of the Abelian Higgs model, the vacuum degeneracy along the flat moduli space can be lifted by coupling the superfield $U$ to a new gauge singlet superfield $\phi'$ via a superpotential term of the form
\begin{equation}\label{PSU2}
P = \frac{1}{8}\,\text{tr}\left[W W\right] + \frac{\lambda'}{2} \phi' \left(\frac{v_u^2}{2} - U^T U\right) \,,
\end{equation}
where $v_u$ is a mass scale that we can choose to be real and positive and which sets the scale of spontaneous  $\text{SU}(2)$ breaking. The new superpotential in Eq.~\eqref{PSU2} now only exhibits $\mathcal{N}=1$ supersymmetry, as the new term breaks $\mathcal{N}=2$ supersymmetry. Next, we derive the equations of motions for the auxiliary fields contained in $V$, $U$, and $\phi'$,
\begin{equation}
D^a =  ig\,\epsilon_{abc}\, U^{b*} U^c \,, \qquad
F_U^{a*} = \lambda' \phi'\, U^a \,, \qquad
F_{\phi'}^* = -\frac{\lambda'}{2} \left(\frac{v_u^2}{2} - U^T U\right) \,.
\end{equation}
Instead of a supersymmetric moduli space, we now find a supersymmetric vacuum at
\begin{equation}
U^a = \frac{v_u}{\sqrt{2}}\,\delta_{a3} \,, \qquad \phi' = 0 \,,
\end{equation}
where the value of $U^TU$ is now fixed and $U^a$ is determined up to an $\text{SU}(2)$ rotation.
Meanwhile, one rotation of the fields $U^a$ still represents an unbroken symmetry, despite the fact that $U^TU$ has nonvanishing expectation value, which means that a $\text{U}(1)$ subgroup of $\text{SU}(2)$ survives in the new ground state after symmetry breaking. The particle spectrum of the theory now consists of: a massless vector multiplet, $V^3$; a charged vector multiplet with mass $m_V = gv_u$ that has ``eaten'' the Goldstone multiplets $U^{1,2}$; and a massive chiral multiplet with mass $m_U = \lambda' v_u/\sqrt{2}$ composed of the multiplets ${U'}^3=U^3-v_u$ and $\phi'$.
 
After symmetry breaking, the theory contains excited states in the form of topologically stable monopoles. To see this, note that the vacuum manifold $\mathcal{M}$ is a 2-sphere $S^2$ spanned by the $\text{SU}(2)$ rotations acting on the fields $U^a$ in the ground state, just like in the nonsupersymmetric case. The vacuum manifold thus has nontrivial homotopy group $\pi_2\left(\mathcal{M}\right) = \mathbb{Z}$, which indicates the existence of monopole solutions. The simplest monopole configuration is the ``hedgehog'' solution~\cite{tHooft:1974kcl,Polyakov:1974ek}, corresponding to radial field profiles of the form
\begin{equation}\label{hedgehog}
U^a = \frac{v_u}{\sqrt{2}} \,f\left(r\right) \frac{x^a}{r} \,, \qquad A^a_0 = 0 \,, \qquad
A^a_i = h\left(r\right) \epsilon_{aij}\,\frac{x^j}{gr^2} \,.
\end{equation}
Now, $r$ denotes the radial coordinate in spherical coordinates rather than cylindrical coordinates, $r=(x^i x^i)^{1/2}$, and the functions $f$ and $h$ are subject to the boundary conditions
\begin{equation}
f\left(0\right) = h\left(0\right) = 0 \,, \qquad f\left(\infty\right) = h\left(\infty\right) = 1 \,.
\end{equation}
From Eq.~\eqref{hedgehog}, we read off that the scalar field profile points into the radial direction, $U^a \propto \hat{\phi^a} \equiv x^a/r$. The same is therefore true for the unbroken symmetry generator. Similarly, one obtains the following gauge-invariant magnetic field strength at large distances,
\begin{equation}
B_i = -\frac{1}{2}\,\hat{\phi}^a\, \epsilon_{ijk}\, F^a_{jk} = \frac{x^i}{gr^3} \,,
\end{equation}
which allows us to identity the magnetic charge of the monopole, $4\pi/g$.
In general, the monopole mass is $2n\,\pi v_u/g$ with $n \in \mathbb{N}$, i.e., the 't Hooft--Polyakov monopole
corresponds to $n=2$.
The mass of the monopole is subject to the Bogomol'nyi bound \cite{Bogomolny:1975de},
\begin{equation}
m_M \geq \frac{4\pi m_V}{g^2} = \frac{4\pi v_u}{g}\,,
\end{equation} 
where the equal sign holds in the Prasad--Sommerfield limit $\lambda'/g \rightarrow 0$~\cite{Prasad:1975kr}. For nonzero values of the ratio of coupling constants, $\lambda'/g$, there exist no analytical expressions for the functions $f$ and $h$ in Eq.~\eqref{hedgehog}. One therefore has to resort to numerical solutions of the field equations, which show that $m_M$ is a monotonically increasing function of the Higgs mass. In the limit $\lambda'/g \rightarrow \infty$, one finds in particular an upper bound $m^{\text{max}}_M \simeq 4\pi m_V/g^2 \times 1.79$~\cite{Kirkman:1981ck,Schellekens:1983yc}.

\subsection{Monopoles and metastable strings}

Let us now combine the constructions in Secs.~\ref{subsec:strings} and \ref{subsec:monopoles} and discuss metastable $B\!-\!L$ strings decaying into short string segments with monopoles and antimonopoles on their ends. Defects of this type form if we embed the electroweak part of the SM gauge group, $G_{\rm EW} = \text{SU}(2)_L\times \text{U}(1)_Y$, in the group $G_{221} = \text{SU}(2)_L\times\text{SU}(2)_R \times \text{U}(1)_{B-L}/\mathbb{Z}_2$ and spontaneously break $G_{221}$ down to $G_{\rm EW}$ in two steps. Hypercharge $Y$ then follows from the linear combination of the neutral $\text{SU}(2)_R$ and $\text{U}(1)_{B-L}$ generators, $Y = T^3_R + \left(B-L\right)/2$. The two symmetry-breaking steps in this model break $\text{SU}(2)_R\times \text{U}(1)_{B-L}/\mathbb{Z}_2$ to $\text{U}(1)_Y$ and end on the vacuum manifold $\mathcal{M} = \text{U}(2)/\text{U}(1) = S^3$. This manifold contains the union of the vacuum manifolds of stable strings and monopoles, $S^1 \cup S^2$, and has trivial homotopy groups $\pi_1\left(\mathcal{M}\right)$ and $\pi_2\left(\mathcal{M}\right)$. The model thus neither features topologically stable monopoles nor strings. Instead, we will see that it can give rise to metastable strings or unstable dumbbells.

In order to break $\text{U}(2) = \text{SU}(2)_R\times \text{U}(1)_{B-L}/\mathbb{Z}_2$ down to $\text{U}(1)_Y$, we shall work with similar Higgs representations as in Secs.~\ref{subsec:strings} and \ref{subsec:monopoles}. Specifically, we introduce a $B\!-\!L$-neutral $\text{SU}(2)_R$ triplet $U$ as well as two oppositely $B\!-\!L$-charged $\text{SU}(2)_R$ doublets $S$, $S_c$,
\begin{equation}
U \sim \left(3,0\right) \,, \qquad S \sim \left(2,q\right) \,, \qquad S_c \sim \left(\bar{2},-q\right) \,,
\end{equation}
under $\text{SU}(2)_R\times \text{U}(1)_{B-L}$. Defects in nonsupersymmetric $\text{U}(2)$ models with triplet and doublet Higgs representations were previously discussed, e.g., in Refs.~\cite{Copeland:1987ht,Kephart:1995cg,Achucarro:1999it,Kibble:2015twa}. In the following, we are, however, interested in the supersymmetric version of the model, which we construct by choosing the K\"ahler potential $K$ and superpotential $P$ as a combination of Eqs.~\eqref{KPU1}, \eqref{KPSU2}, and \eqref{PSU2}, supplemented by an additional mass term in $P$,%
\footnote{Our model is different from standard left-right-symmetric models, as we work with neutral triplets. In left-right-symmetric models, the triplets typically carry $\text{U}(1)$ charge and thus occur in pairs~\cite{Aulakh:1998nn,Babu:2008ep}.}
\begin{equation}\label{KPU2}
\begin{split}
  K & = U^\dagger e^{2gV} U + S^\dagger e^{2\left(g\tilde{V} + g'qV'\right)}S + S_c^\dagger e^{-2\left(g\tilde{V}+g'qV'\right)}S_c + \phi^\dagger\phi +
  {\phi'}^{\dagger}\phi' \,, \\
  P & = \frac{1}{8}\, \text{tr}\left[W W\right] + \frac{1}{4}\, W'W' + 2 h\, S^T_c\,\tilde{U}\,S \\
  &\: + \frac{\lambda'}{2} \phi' \left(\frac{v_u^2}{2} - U^T U\right) + \lambda\phi\left(\frac{v_s^2}{2} - S^T_cS\right) - h v_u \,S^T_c S \,.
\end{split}
\end{equation}
Here, $U = \left(U^1,..,U^3\right)^T$ is the triplet field written as a vector in the triplet representation; $\tilde{U} = U^a \tau^a/2$ is the triplet field written as a matrix in the doublet representation; $V = V^a T^a$ is the $\text{SU}(2)_R$ vector field in the triplet representation; $\tilde{V} = V^a \tau^a/2 \equiv T_R$ is the $\text{SU}(2)_R$ vector field in the doublet representation; $V'$ is the $\text{U}(1)_{B-L}$ vector field; and $W$ and $W'$ are the supersymmetric $\text{SU}(2)_R$ and $\text{U}(1)_{B-L}$ field strengths.
The covariant derivative in the bosonic sector, induced by the K\"ahler potential, is given by $D_\mu S = \partial_\mu S + i(g\tilde{V} + g'qV')S$.

The terms involving the Yukawa coupling $h$ are introduced for the following reason: First, the trilinear coupling, coupling the fundamental triplet $U^a$ to the composite triplet $S^T_c\tau^a S$ ensures that a $\text{U}(1)$ subgroup survives after symmetry breaking. Without this term, the initial $\text{U}(2)$ group would in general be broken completely, which in our case would mean that no hypercharge gauge group $\text{U}(1)_Y$ would remain in the electroweak sector. Second, the mass term for the pair of doublet fields, i.e., the last term in $P$ in Eq.~\eqref{KPU2}, ensures that $\text{U}(2)$ symmetry breaking results in a supersymmetric vacuum with $\langle P\rangle = 0$. This serves the purpose to separate the energy scales of $\text{SU}(2)_R$  and $\text{U}(1)_{B-L}$ breaking from the energy scale of supersymmetry breaking. Without this mass term, we would generically expect a contribution to the gravitino mass from the $\text{U}(2)$ sector of the order of $\langle P\rangle/M^2_{\text{P}}  \sim hv_uv_s^2/M^2_{\text{P}}$. However, if $\langle P\rangle = 0$ after $\text{U}(2)$ symmetry breaking, we retain the possibility that a separate supersymmetry-breaking sector results in a hierarchically smaller gravitino mass.

As in Secs.~\ref{subsec:strings} and \ref{subsec:monopoles}, the model exhibits again $D$-flat directions that are lifted by the coupling to the singlets $\phi$ and $\phi'$. The supersymmetric true vacuum then corresponds to
\begin{equation}
U^a = \frac{v_u}{\sqrt{2}}\,\delta_{a3} \,, \qquad S = S_c =  \frac{v_s}{\sqrt{2}} \begin{pmatrix} 1 \\ 0 \end{pmatrix} \,, \qquad \phi' = \frac{hv_s^2}{\sqrt{2}\lambda' v_u} \,, \qquad \phi = 0 \,.
\end{equation}
The fundamental triplet $U^a$ and the composite triplet $S^T_c\tau^aS$ are parallel in this vacuum configuration, as desired, thanks to the Yukawa coupling $h \neq 0$ in Eq.~\eqref{KPU2}. As mentioned above, this is necessary to keep an unbroken $\text{U}(1)$ symmetry in the vacuum. Without the Yukawa coupling $h$, the relative orientation of $U^a$ and $S^T_c\tau^aS$ would not be fixed.                                         

Next, let us discuss the mass spectrum of the model. In order to identify the mass eigenstates, we must shift the chiral multiplets around their vacuum expectation values,
\begin{equation}
U^3 = \frac{v_u}{\sqrt{2}} + U^{3\prime} \,, \qquad S = \begin{pmatrix} \frac{v_s}{\sqrt{2}} + S^{0\prime} \\ S^- \end{pmatrix} \,, \qquad S_c = \begin{pmatrix} \frac{v_s}{\sqrt{2}} + S^{0\prime}_c \\ S^+ \end{pmatrix} \,, \qquad \phi' = \frac{hv_s^2}{\sqrt{2}\lambda'v_u} + \hat{\phi} \,.
\end{equation}
Then, by inspecting the terms linear in the vector fields $V$, $\tilde{V}$, and $V'$ in the K\"ahler potential in Eq.~\eqref{KPU2}, we can identify the Goldstone multiplets,
\begin{equation}
\Pi^\mp = \frac{1}{\sqrt{2v_u^2 + v_s^2}} \left(\sqrt{2}v_u\, U^\mp + v_s\,S^\mp\right) \,, \qquad \Pi^0 = \frac{1}{\sqrt{2}}\left(S^{0\prime} - S^{0\prime}_c\right) \,,
\end{equation}
where $U^\pm = \left(U^1 \mp i U^2\right)/\sqrt{2}$, which are respectively ``eaten'' by the vector multiplets
\begin{equation}
  V^\pm = \frac{1}{\sqrt{2}}\left(V^1 \mp iV^2\right)\,, \qquad
  V_X = \left(\cos\Theta\, V^3 + \sin\Theta\, V'\right) \,,
\end{equation}
with $\tan\Theta = 2g'q/g$. The vector multiplet orthogonal to these two fields,
\begin{equation}
V_Y = -\sin\Theta\, V^3 + \cos\Theta\, V' \,, 
\end{equation}
remains massless, while the vector multiplets $V^\pm$ and $V_X$ acquire masses 
\begin{equation}
\label{mVmX}
m^2_V = \frac{g^2}{2} \left(2v_u^2 + v_s^2\right) \,, \qquad
m^2_X = \frac{g^2}{2\cos^2\Theta} v_s^2 \,.
\end{equation}
In addition to the Goldstone multiplets $\Pi^\pm$ and $\Pi^0$ and vector multiplets $V^\pm$, $V_X$, and $V_Y$, we are left with six chiral multiplets, $\Sigma^\pm$, $\Sigma^0$, $U^{3\prime}$, $\phi$, and $\hat{\phi}$. The mass matrix of these fields follows from the quadratic part of the superpotential,
\begin{equation}
\label{eq:Pm}
P_m = - \sqrt{2}h\left(\frac{2v_u^2+v_s^2}{v_u}\right) \Sigma^-\Sigma^+ 
- \lambda' \frac{v_u}{\sqrt{2}} \hat{\phi}\, U^{3\prime} - v_s\left(\lambda \phi - h U^{3\prime}\right) \Sigma^0
- \frac{h v_s^2}{2\sqrt{2}v_u}\, U^{3\prime}U^{3\prime} \,,
\end{equation}
where the linear combinations
\begin{equation}
\Sigma^\pm = \frac{1}{\sqrt{2v_u^2 + v_s^2}}\left(-v_s\, U^\pm + \sqrt{2}v_u\, S^\pm\right) \,, \qquad \Sigma^0 = \frac{1}{\sqrt{2}}\left(S^{0\prime} + S^{0\prime}_c\right) \,,
\end{equation}
are orthogonal to the Goldstone multiples $\Pi^\pm$ and $\Pi^0$, respectively. We emphasize again the role of the Yukawa coupling: in absence of the $h$-dependent terms in Eq.~\eqref{eq:Pm}, the superpotential $P_m$ simply corresponds to the mass terms discussed in Secs.~\ref{subsec:strings} and \ref{subsec:monopoles}, i.e., the mass terms for $\text{SU}(2)_R$
and $\text{U}(1)_{B-L}$ breaking in isolation.

For suitably chosen parameter values, the model discussed in this section allows us to break $\text{SU}(2)_R \times \text{U}(1)_{B-L}$ down to $\text{U}(1)_Y$ in two subsequent steps, each of which corresponding to a cosmological phase transition in the early universe. In the first step, a nonvanishing triplet expectation value $\langle U^a\rangle$ breaks $\text{SU}(2)_R$ to $\text{U}(1)_R$; and then in a second step, nonvanishing doublet expectation values $\langle S\rangle$ and $\langle S_c\rangle$ break $\text{U}(1)_R\times \text{U}(1)_{B-L}$ down to $\text{U}(1)_Y$. Note that analogous symmetries are present in the electroweak sector, where $\text{SU}(2)_L$ contains the subgroup $\text{U}(1)_L$ and where $\text{U}(1)_L \times \text{U}(1)_Y$ contain in turn the electromagnetic subgroup $\text{U}(1)_Q$; even though electroweak symmetry breaking does not involve any Higgs triplets.

Up to now, we treated the $\text{U}(1)_{B-L}$ gauge coupling times the charge of the doublet fields, $qg'$, as a free parameter. This is no longer possible as soon as one begins to consider embeddings of our model in either of the symmetry breaking chains in Eqs.~\eqref{embSU5} and \eqref{embLR}.
  In Pati--Salam or $\text{SO}(10)$ GUT extensions of the SM, the Higgs doublets $S$ and $S_c$ are embedded into Pati--Salam $\left(\mathbf{4},1,\mathbf{2}\right) \sim \chi_L$ and $\left(\mathbf{\bar{4}},1,\mathbf{\bar{2}}\right) \sim \chi_R^c$ representations, or into $\mathbf{16}$, $\mathbf{\overline{16}}$ representations of $\text{SO}(10)$, respectively [see~Eq.~\eqref{SU5B}]. Here, the Higgs doublets $S$,
 $S_c$ are identified as the ``lepton doublets'' in $\chi_L$, $\chi^c_R$. 
For the Pati--Salam embedding, the
covariant derivative in the bosonic sector reads
$D_\mu\chi_L  = \partial_\mu \chi_L + i\left(g T^a_RV^a + g'\frac{1}{2}\left(B\!-\!L\right) V'\right)\chi_L$. The normalization
condition $g^{\prime 2}/4\,\text{tr}\left[(B\!-\!L)^2\right] = g^2\text{tr}\left[(T^3_R)^2\right]$ 
then implies $g'\sqrt{2/3} = g$, which corresponds to the mixing angle
$\tan\Theta = -\sqrt{3/2}$. The covariant derivative with the two $\text{U}(1)$ factors
$\text{U}(1)_R$ and $\text{U}(1)_{B-L}$ then reads 
$D_\mu \chi_L = \partial_\mu \chi_L + ig\,\big(T^3_R V^3+ \sqrt{3/2}~\frac{1}{2}\left(B\!-\!L\right) V'\big)\, \chi_L$.
Note that the field $S$ carries charge $\pm 1/2$ with respect to the generators $T^3_R$ and $\frac{1}{2}(B - L)$,
respectively.

Embedding the doublets $S$, $S_c$ in $\mathbf{16}$-, $\mathbf{\overline{16}}$-plets
  $\Phi$, $\Phi^c$ of $\text{SO}(10)$, as in Eq.~\eqref{SU5B},
  implies that heavy Majorana neutrino masses must
be generated by the nonrenormalizable operator
\begin{equation}
  \mathcal{L}_n = \frac{1}{M_*}\,h_{ij}\, S^T L^c_i S^T L^c_j \subset
\frac{1}{M_*}\, h_{ij}\, \Phi^c \psi_i {\Phi^c} \psi_j \,.
\end{equation}
  Here, the fields $L^c_i = (n^c_i,e^c_i)^T$,
  $i=1,..,3$, denote the $\text{SU}(2)_R$ doublets of right-handed neutral and charged
  leptons that are contained in the $\text{SO}(10)$ $\mathbf{16}$ representations
  $\psi_i$ of matter,
  and $h_{ij}$ are Yukawa couplings. Alternatively, one can follow
  Eq.~\eqref{SU5A} and break $\text{SO}(10)$ with $\mathbf{126}$-,
  $\mathbf{\overline{126}}$-plets $\tilde{\Phi}$, $\tilde{\Phi}^c$ containing
  the $\text{SU}(5)$ singlets $\tilde{S}$, $\tilde{S_c}$. Heavy
  neutrino masses are now generated by the renormalizable couplings
  \begin{equation}
  \mathcal{L}_n = h_{ij} \tilde{S} n^c_i n^c_j \subset
  h_{ij} \tilde{\Phi} \psi_i \psi_j \ ,
\end{equation}
as assumed, e.g., in Ref.~\cite{Buchmuller:2012wn}.
The VEVs of $\tilde{S}$, $\tilde{S_c}$ leave a $\mathbb{Z}_2$ discrete 
symmetry unbroken, which leads to topologically stable strings.

The cosmological realization of the two symmetry-breaking stages in our model (in the form of cosmological phase transitions) leads to the formation of defects: monopoles in the first step and strings in the second step. For a monopole--string--antimonopole
configuration, the magnetic fluxes of the string and the (anti)monopole have to match\footnote{Note that, in the $\text{U}(2)$ model, this is only possible if $g'q$ and $g$ are integer multiples of each other. This is guaranteed by the Pati--Salam embedding.} (see, e.g., Ref.~\cite{Kibble:2015twa}). The string solution with lowest energy has winding number $n=1$.
As the symmetry breaking field $S$ has charge $1/2$, it carries magnetic
flux $4\pi/g$. This can be matched by a $n=2$ monopole with mass $m_M \sim
4\pi v_u/g$. Together with the string tension $\mu \simeq 2\pi v_s^2$, we then
obtain for the parameter $\kappa$, which controls the metastability of
cosmic strings, 
\begin{equation}\label{kappa}
  \kappa = \frac{m_M^2}{\mu} \sim \frac{8\pi}{g^2}\frac{v_u^2}{v_s^2} \,.
\end{equation}
In supersymmetric theories, one expects $g^2 \sim 1/2$ at the unification
scale. This implies $\sqrt{\kappa} \sim 7\,v_u/v_s$.
As we shall see in the following section, metastable strings can be
relevant for GWs in the PTA band for $\sqrt{\kappa} \gtrsim 8$, which
corresponds to $v_u \gtrsim v_s$.
Note that the model predicts confined as well as unconfined magnetic flux
for the monopole, which is given by $(4\pi/g) \sin^2\Theta$ (for a discussion,
see Ref.~\cite{Kibble:2015twa}).

An important open question concerns the range of validity of the
relation in Eq.~\eqref{kappa}. The size of the magnetic cores of the monopole and the string are given by
$m_V^{-1}$ and $m_X^{-1}$, respectively. The string decay rate will also
be affected by
the false vacuum cores, whose size is given by
the Higgs masses $m_U$ and $m_S$ for monopole and string, respectively.
Moreover, Eq.~\eqref{kappa} uses estimates for the mass and tension of
an isolated monopole 
as well as an isolated string, respectively. So far, no calculations have carried out for
spatially extended composite defects.
As $v_u$ approaches zero, 
the semiclassical approximation used in the derivation of
Eq.~\eqref{kappa} breaks down, and metastable strings turn into
dumbbells that decay immediately.

In the case where $\text{SU}(2)_R\times\text{U}(1)_{B-L}$ is broken
to $\text{U}(1)_Y$ by VEVs of the doublets $S$ and $S_c$ only,
dumbbells or
X-strings form, which are completely analogous to the Z-strings
of the SM \cite{Nambu:1977ag}.
For $|\tan{\Theta}| = \sqrt{3/2}$,
X-strings are known to be unstable
\cite{Vachaspati:1992fi,Achucarro:1999it,Kanda:2022xrz,Kanda:2023yyz}.

Metastable strings are a generic feature of GUTs. The supersymmetric breaking of $\text{U}(2) = \text{SU}(2)_{R}\times\text{U}(1)_{B-L} /\mathbb{Z}_2$ is the simplest example (albeit a representative one) of a much richer structure that occurs in realistic GUTs. It is an intriguing prospect that the metastability of cosmic strings might be tested with gravitational waves, which would provide  direct information about the energy scales of GUT symmetry breaking stages.

\section{Stochastic gravitational-wave background}
\label{sec:sgwb}

The SGWB sourced by a metastable cosmic-string network was recently computed in Ref.~\cite{Buchmuller:2021mbb} (see also Refs.~\cite{Gouttenoire:2019kij,Dunsky:2021tih}). As for stable cosmic strings, a population of string loops with number density $\overset{\circ}{n}(\ell, t')$ radiates GWs with the power density per frequency~\cite{Blanco-Pillado:2017oxo,Auclair:2019wcv}
\begin{align}
 P_\text{gw}(t', f') = G \mu^2 \sum_{k = 1}^{k_\text{max}} \frac{\ell}{f'}\,\overset{\circ}{n}\left(\ell, t'\right) P_k \,,
\end{align}
where $f' = 2 k/\ell$ indicates the GW frequency, emitted by a loop of length $\ell$ oscillating in its $k$th harmonic excitation; $t'$ is the time of GW emission and $P_k = \Gamma/(k^{4/3} \zeta[4/3])$ with $\Gamma \simeq 50$ is the power emitted by a single loop (assuming the emission is dominated by the contribution from cusps). Integrating over $t'$, we obtain the spectral energy density in GWs today normalized by the critical energy density,
\begin{align}
 \Omega_\text{gw}(t_0, f) = \frac{16 \pi (G\mu)^2}{3 H_0^2 f} \sum_k k P_k \int_0^{z_i} \frac{dz'}{H(z') (1 + z')^6}\, \overset{\circ}{n}(2 k/f', t(z')) \,.
 \label{eq:OmegaGW}
\end{align}
Here $H(z)$ is the Hubble parameter, we have switched the time-variable to redshift $z$, and the argument of the loop number density ensures that we are accounting for all GWs emitted at frequency $f'$ such that after red-shifting, they are observed at frequency $f$ today.

The remaining challenge is to determine the loop number density $\overset{\circ}{n}(\ell, t')$. In the limit of stable cosmic strings, we adopt the velocity-dependent one-scale (VOS) model~\cite{Martins:1996jp,Martins:2000cs} within the Nambu--Goto framework, in which one-dimensional string loops are formed at a fixed fraction $\alpha$ of the horizon and then shrink due to GW emission, $\ell(t) = \alpha t' - \Gamma G \mu (t - t')$. In this case, the loop number density can be determined analytically by solving the corresponding kinetic equation, up to integration constants which can be extracted from simulations~\cite{Blanco-Pillado:2013qja}. For example, in a radiation-dominated background, this yields
\begin{align}
 \overset{\circ}{n}^\text{rad}_\infty(\ell, t) = \frac{B}{t^{3/2} (\ell + \Gamma G \mu t)^{5/2}} \, \Theta(\alpha t - \ell) \,,
 \label{eq:nstable}
\end{align} 
where $B = 0.18$ and $\alpha = 0.1$ are obtained from a fit to numerical simulations. The subscript $\infty$ (for $\kappa \rightarrow \infty$) refers to stable cosmic strings.

For metastable cosmic strings, the kinetic equations are modified to take into account the decay of string loops to segments through the formation of a monopole--antimonopole pair, as well as the formation of segments from longer segments and super-horizon strings~\cite{Buchmuller:2021mbb}. If the monopoles carry no unconfined flux, the segments themselves can have cosmological lifetimes and contribute to the SGWB~\cite{Martin:1996cp,Leblond:2009fq}. However, as demonstrated in Ref.~\cite{Buchmuller:2021mbb}, the GW spectrum generated by cosmic string loops alone provides a good approximation to the full spectrum in most of the parameter space, even when there is a contribution from segments.
Moreover, the example in Sec.~\ref{sec:metastrings} has unconfined flux. We therefore focus
  on the GW spectrum from string loops, in which case the key change compared to stable cosmic strings is an additional decay term in the kinetic equation for the loop number density accounting for the monopole--antimonopole formation on the loops. Matching the number density to Eq.~\eqref{eq:nstable} at early times, $t \ll t_s =  1/\Gamma_d^{1/2}$, then yields for the loop number density of the metastable cosmic string network at $t > 1/\Gamma_d^{1/2}$,
\begin{align}
 \overset{\circ}{n}^\text{rad}(\ell, t) = \frac{B}{t^{3/2} (\ell + \Gamma G \mu t)^{5/2}}\, e^{- \Gamma_d [ \ell (t - t_s) + \tfrac{1}{2} \Gamma G \mu (t - t_s)^2]} \, \Theta(\alpha t_s - \ell - \Gamma \mu ( t - t_s)) \,.
 \label{eq:nmetastable}
\end{align} 
Here, the exponential factor accounts for the decay of the loops at $t > t_s$ through the generation of monopoles, and the Heaviside function ensures that loop formation only occurs at $t < t_s$. For expressions for the loop number densities involving evolution during the matter-dominated era as well as expressions for the number densities of super-horizon strings and segments, see Ref.~\cite{Buchmuller:2021mbb}. 

\begin{figure}[t]
\centering
\includegraphics[width = 0.6 \textwidth]{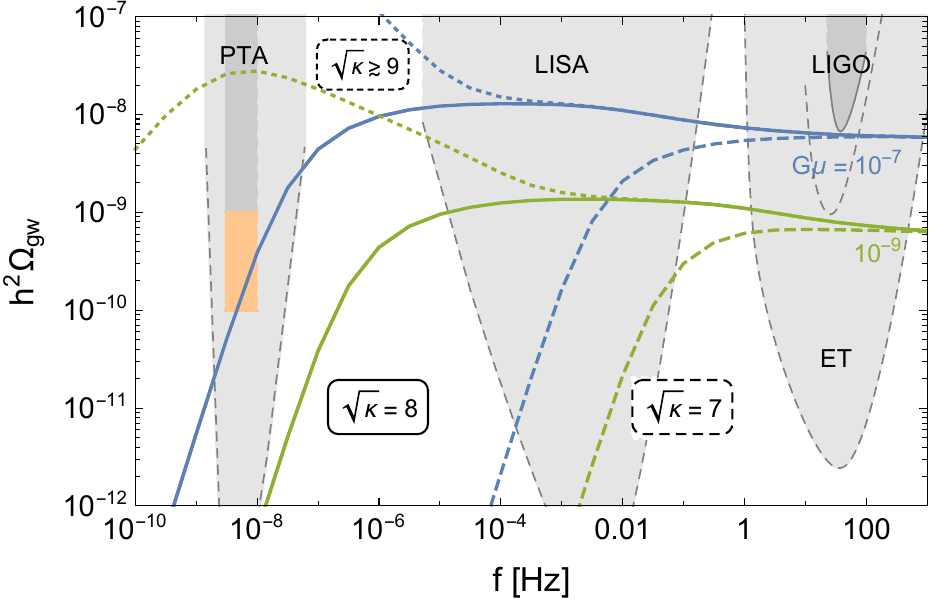}
\caption{GW spectrum from metastable cosmic strings with a string tension of $G\mu = 10^{-7}$ (blue) and $G\mu = 10^{-9}$ (green). Different line styles indicate different string lifetimes, ranging from $\sqrt{\kappa} = 7$ (dashed) over $\sqrt{\kappa} = 8$ (solid) to the limit of (quasi-)stable strings $\sqrt{\kappa} \gtrsim 9$ (dotted).
The gray-shaded areas indicate the sensitivity of current (solid) and planned (dashed) GW experiments. The orange region indicates the preferred region of the possible SGWB signal observed by PTAs.}
\label{fig:spectrum}
\end{figure}

Fig.~\ref{fig:spectrum} shows the GW spectrum obtained by inserting Eq.~\eqref{eq:nmetastable} (and corresponding expressions for the matter-dominated era) into Eq.~\eqref{eq:OmegaGW}. The dotted black curves show the limit of stable cosmic strings, $\kappa \rightarrow \infty$, whereas the colored curves show the prediction for the spectrum for two different values of the ratio of the symmetry breaking scales $\kappa$ and the string tension $\mu$. Large frequencies correspond to GWs produced at early times, and hence the spectrum produced by stable and metastable strings is identical, featuring a plateau at
\begin{align}
 \Omega_\text{gw}^\text{plateau} \simeq \frac{128 \pi}{9} B \, \Omega_r \left(\frac{G \mu}{\Gamma} \right)^{1/2} \,,
\end{align}
where $\Omega_r h^2 = 4.15 \cdot 10^{-5}$ is the density parameter of radiation today.
At lower frequencies, the earlier decay of the metastable cosmic
string loops suppresses the GW signal, leading to a drop in the
spectrum proportional to $f^2$. This drop sets in at a frequency~\cite{Buchmuller:2021mbb}
\begin{equation}
\label{flow}
f_\text{low} \sim 3\cdot 10^{-9}\,\textrm{Hz}
\left(\frac{50}{\Gamma}\right)^{3/4}
\left(\frac{10^{-8}}{G\mu}\right)^{1/2}
\exp{\left(-\pi\left(\frac{\kappa}{4}-16\right)\right)} \ .
\end{equation}
GWs from metastable strings can be observed if their decay happens
sufficiently late such that their redshifted frequencies are not much below
 $f_\text{low}$. On the observational side, a lower cutoff on the
measurable frequencies at PTAs is given by the observation times, i.e.,
$f^\text{PTA} \sim (10~\text{yr})^{-1} \simeq 3~\text{nHz}$.
The cutoff $f_\text{low}$ depends exponentially on $\kappa$, and
from Eq.~\eqref{flow} one reads off that the condition for a potential
discovery of GWs at PTAs,  $f_{\text{low}}  \lesssim f^\text{PTA}$,
requires $\sqrt{\kappa} \gtrsim 8$.
Note that the predicted $f^2$ spectrum is remarkably different from the $f^3$ scaling expected for causal GW sources (including, e.g., GWs from domain walls) and is due to the fact that the cosmic-string network acts as a GW source over time scales much larger than a Hubble time.

PTA collaborations across the world have recently reported the observation of a stochastic common-spectrum process\cite{Arzoumanian:2020vkk,Goncharov:2021oub,Chen:2021rqp,Antoniadis:2022pcn} showing evidence for Hellings--Downs spatial correlations~\cite{NANOGrav:2023gor,Antoniadis:2023ott,Xu:2023wog,Reardon:2023gzh}, the hallmark signature of a SGWB. While the most plausible source remains supermassive black-hole binaries, the observed tension with common astrophysical population models~\cite{NANOGrav:2023hfp,Antoniadis:2023xlr} motivates a thorough investigation of a possible cosmological contribution. Upcoming data will improve our understanding of the spectral tilt, the isotropy and the presence of resolvable individual sources in this SGWB, which will all help to distinguish an astrophysical from a cosmological origin.
With these promises and caveats in mind, we now focus in more detail
on the GW signal of metastable cosmic strings in the PTA frequency
band. As shown in Refs.~\cite{Buchmuller:2020lbh,Buchmuller:2021mbb,Dunsky:2021tih}, this
could explain the observed GW signal for $10^{-11} \lesssim G \mu \lesssim
10^{-7}$ and $\sqrt{\kappa} \gtrsim 8$. For models of hybrid and
tribrid inflation compatible with these values, see,
e.g., Refs.~\cite{Buchmuller:2019gfy,Masoud:2021prr,Afzal:2022vjx}. From Eq.~\eqref{kappa},
one reads off that $\sqrt{\kappa} \simeq 8$ can be achieved if the two symmetry breaking scales $v_u$
and $v_s$ are close to each other. 
Metastable cosmic strings thus lead to observable effects at PTAs for $v_u \gtrsim
v_s$. If the astrophysical origin of the currently observed signal
should be confirmed, which corresponds to interpreting the current PTA data as an upper
bound on a cosmological signal, this would shift the interest to the region $v_u \lesssim v_s$ (or to smaller values of $G\mu$),
which remains compatible with such a constraint while still allowing for a large SGWB signal in the LISA and LIGO bands.%
\footnote{The described connection between PTA observation times
  and GUT-scale parameters may appear surprising, but an analogous case
  is known from neutrino physics. Neutrino mass differences 
$\sqrt{\Delta m^2} \sim 0.05$~eV could only be discovered in atmospheric neutrino oscillations
because the oscillation length is $L \sim 10^4$~km (earth diameter)~\cite{ParticleDataGroup:2022pth}. Smaller mass differences, 
e.g. $\sqrt{\Delta m^2} \sim 0.005$~eV, would have remained unobserved. \vspace{2mm}}
\begin{figure}[t]
\centering
\includegraphics[width = 0.49 \textwidth]{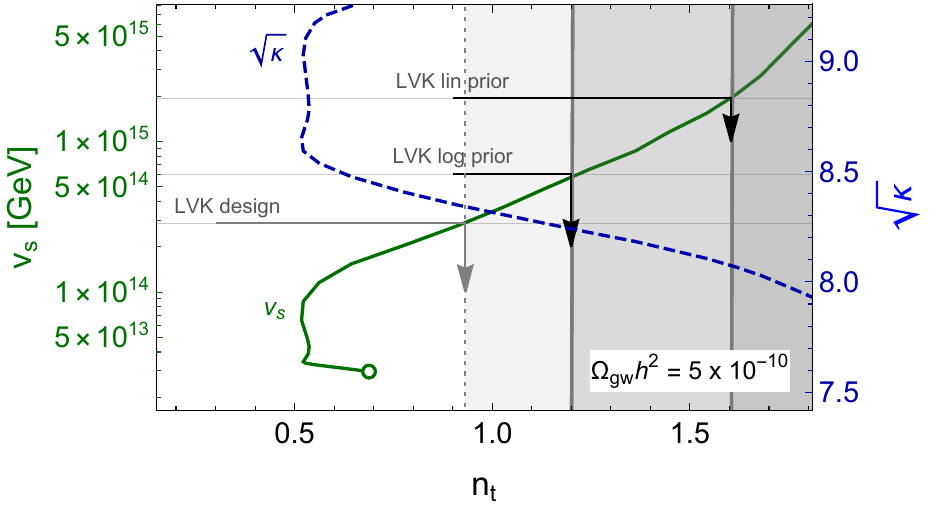} \hfill
\includegraphics[width = 0.49 \textwidth]{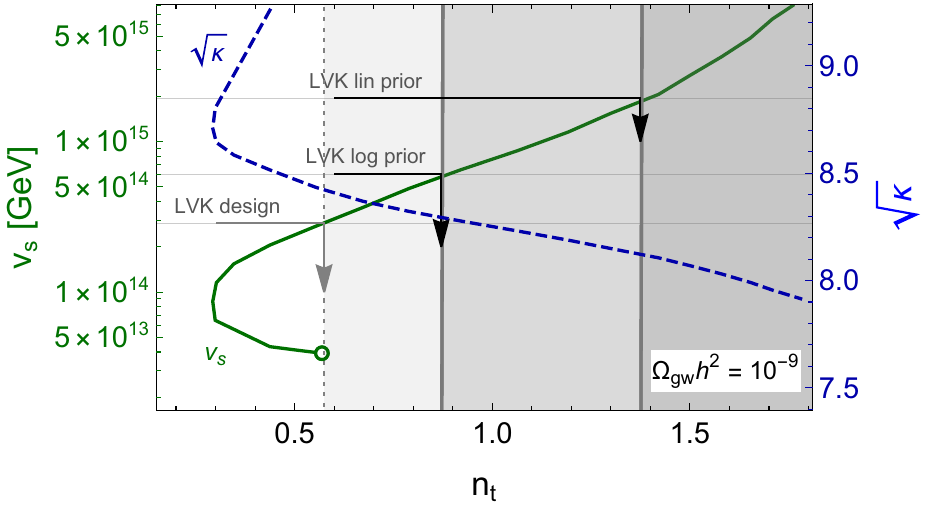}
\caption{$U(1)$ symmetry breaking scale $v_s$ (solid green, left axis) and ratio $\kappa^{1/2}$ between monopole mass and string tension (dashed blue, right axis) as a function of the predicted spectral tilt $n_t$ of the SGWB, assuming an amplitude $\Omega_\text{gw}^\text{PTA} h^2 \simeq 5 \cdot 10^{-10}$ (left) and  $\Omega_\text{gw}^\text{PTA} h^2 \simeq 10^{-9}$ (right) at $f = 3$~nHz. The grey region indicates values $G\mu > 1.5 \cdot 10^{-8}$ ($G\mu > 1.6 \cdot 10^{-7}$) which are disfavoured by LIGO-Virgo--KAGRA (LVK) for a logarithmic (linear) prior on the amplitude of the SGWB signal in the LVK band~\cite{KAGRA:2021kbb}. 
}
\label{fig:nt}
\end{figure}

Within the approximations mentioned above, the GW spectrum from metastable cosmic strings only depends on two parameters, the string monopole mass $m_M$ and the string tension $\mu$, or, dropping the logarithmic dependence on the Yukawa couplings, the two symmetry breaking scales $v_u$ and $v_s$. Interpreting the PTA data as a SGWB, the current data~\cite{Reardon:2023gzh,NANOGrav:2023gor,Antoniadis:2023ott} indicate an amplitude of $10^{-10} \lesssim \Omega_\text{gw}^\text{PTA} h^2 \lesssim 10^{-9}$ at the PTA peak sensitivity $f_\text{PTA} = 3$~nHz. The spectral index $n_t = d\ln\Omega_\text{gw}/d\ln f$ is less constrained and varies more significantly across the different data sets, $0 \lesssim n_t \lesssim 3$, with the PPTA data set preferring slightly smaller values, the EPTA 10.5 year data set preferring larger values and the NANOGrav data lying in the middle. Upcoming data and analysis will significantly improve the measurement of the spectral tilt, allowing a distinction between different SGWB sources. Remarkably, within the framework of metastable cosmic strings, the two observables $\Omega_\text{gw}^\text{PTA}$ and $n_t$ allow to determine the two model parameters, $v_u$ and $v_s$.

This is shown in Fig.~\ref{fig:nt}, where we fix the amplitude of the GW spectrum at $f = 3$~nHz to two distinct reference values, $\Omega_\text{gw}^\text{PTA} h^2 = 5 \cdot 10^{-10}$ (left) and  $\Omega_\text{gw}^\text{PTA} h^2 =10^{- 9}$ (right),  and show how a future improved measurement of $n_t$ (x-axis) would determine the GUT-scale symmetry breaking parameters (on the two vertical axes).%
\footnote{Since the cosmic string signal is not a perfect power law over the frequency range of PTAs, the precise value of the tilt $n_t$ (both in the model prediction and signal reconstruction) depends on the underlying assumptions. Here, for concreteness, we determine $n_t$ by linearly interpolating between the signal predictions at 2 and 4~nHz. Note that $n_t$ is related to the often quoted tilt $\alpha$ of the dimensionless characteristic strain and to the spectral index ($-\gamma$) of the timing-residual power spectral density as $n_t = 2\alpha + 2 = 5 - \gamma$.}
The limit of stable cosmic strings requires $G\mu \simeq 4 \cdot 10^{-11}$ ($ 7 \cdot 10^{-11}$) yielding $n_t \simeq 0.7$ ($0.6$) to reproduce this SGWB amplitude for $\Omega_\text{gw}^\text{PTA} h^2 = 5 \cdot 10^{-10}$ ($10^{-9}$).
As the cosmic-string lifetime and hence $\kappa$ is reduced, the string tension $\mu$ needs to be increased to maintain the same SGWB amplitude at 3~nHz. For quasi-stable strings, an increase in $G\mu$ comes with a decrease in $n_t$, until with a further decrease of $\kappa$ the $f^2$ part of the spectrum enters the PTA band and the spectral index starts increasing again. For large string tensions (large $v_s$), the desired SGWB amplitude can only be achieved by significantly reducing the string lifetime (reducing $\kappa$), recovering the asymptotic $f^2$ scaling. Of course in this case, unless the reheating temperature is very low or a non-standard cosmological history is invoked~\cite{Gouttenoire:2019kij,Allahverdi:2020bys} (see, e.g., Ref.~\cite{Lazarides:2021uxv}), the SGWB will exceed the bound $\Omega_\text{gw} \leq 5.8 \cdot 10^{-9}$ ($1.7 \cdot 10^{-8}$ for a linear prior) set by the LIGO--Virgo--KAGRA (LVK) collaboration in the 100~Hz range~\cite{KAGRA:2021kbb}. If the current preference for a spectral index larger than $n_t \simeq 1$ persists, this moreover disfavours the limit of stable strings resulting in a sweet spot with $v_s = \text{ few } \times 10^{14}$~GeV and $\sqrt{\kappa} \simeq 8.3$ (see also Ref.~\cite{NANOGrav:2023hvm}).

We conclude this section by drawing attention to some theoretical uncertainties and open questions in the calculation of the GW spectrum. Our calculations here are based on the Nambu--Goto action, taking cosmic strings to be infinitely thin, and moreover we focus on the GW emission by cusps on the cosmic string loops. Alternatively, cosmic strings can be modeled using lattice simulations of classical field theory (Abelian Higgs model) and one may consider GW emission from kinks as well as GW bursts. For reviews and more detailed discussions, see Refs.~\cite{Auclair:2019wcv,Blanco-Pillado:2023sap}. In summary, a robust understanding of the substructure on string loops remains challenging but is crucial to accurately estimate the GW spectrum.

\section{Stable and quasi-stable strings}
\label{sec:qscs}

The GW spectrum of metastable strings in the PTA frequency band and above
agrees with the one of stable strings for monopole-mass--string-tension ratios
$\sqrt{\kappa} \gtrsim 9$ \cite{Buchmuller:2021mbb}.
These strings are
usually referred to as quasi-stable. 
They can explain the PTA results for $G\mu = 10^{-(11\cdots10)}$, corresponding to a spectral tilt of $0 \lesssim n_t \lesssim 0.8$~\cite{Ellis:2020ena,Blasi:2020mfx,Bian:2020urb,Blanco-Pillado:2021ygr,EPTA:2023hof}, where the upper bound is relatively sensitive to the details of the modelling of the cosmic-string network~\cite{Blanco-Pillado:2021ygr}.

Many studies have explored possible connections between an
intermediate string scale $v_s \sim 10^{13}~\text{GeV}$ 
and other predictions of
supersymmetric and nonsupersymmetric GUT models. In
nonsupersymmetric $\text{SO}(10)$ models, requiring gauge coupling
unification together with an intermediate string scale significantly
restricts the allowed symmetry breaking chains. Moreover, the
unification scale has to be large enough such that the proton decay
rate is smaller than current experimental bounds \cite{King:2020hyd,King:2021gmj}.
The situation is similar in supersymmetric $\text{SO}(10)$ models
\cite{Chigusa:2020rks}. Since $\text{SO}(10)$ GUTs with a large string
scale around $10^{13}~\text{GeV}$ generically contain heavy Majorana neutrinos,
weakly interacting neutrinos with sub-$\text{eV}$ Majorana masses, as
well as leptogenesis are naturally incorporated. The discussion of 
$\text{SO}(10)$ models can be extended to $\text{E}_6$ models where
new types of monopoles and strings appear \cite{Chakrabortty:2020otp}.

One may also consider extensions of $\text{SO}(10)$ models with a
Peccei--Quinn symmetry whose breaking yields an axion
\cite{Lazarides:2022ezc}. The model predicts two types of monopoles,
related to the GUT scale and an intermediate scale, and in addition
topologically stable strings produced at an intermediate scale below
$10^{13}~\text{GeV}$. The related SGWB is too weak to be observed by
PTAs or LVK, but can be detected at SKA, LISA and ET/CE. On the other hand, for baryogenesis after
primordial-back-hole evaporation as discussed in Ref.~\cite{Datta:2020bht}, the predicted scale of the $B\!-\!L$ strings  is too large to be consistent with
PTA data. These are some examples of the discriminating power of SGWB signals in the PTA band.

So far, we have assumed that a fundamental GUT is broken to the
SM by a sequence of symmetry breakings that are all
realized by the Higgs mechanism. One can then expect a plethora of
topological defects produced in cosmological phase
transitions as sources of a SGWB. However, it is far from obvious that
all symmetry breakings are realized by the Higgs mechanism.
Nonsupersymmetric GUTs
suffer from severe fine-tuning problems, and supersymmetric GUTs with
realistic fermion mass matrices require large Higgs representations,
which make them almost intractable. Attractive alternatives are GUTs
in higher dimensions and in string theories. Compactification to four
dimensions will then reduce the GUT group to a subgroup whose
further breaking to the SM group could then proceed via the
Higgs mechanism (for a review and references, see, e.g., Ref.~\cite{Raby:2017ucc}).
This still leaves room for some topological
defects. Clearly, the discovery of monopoles or evidence for strings
in a SGWB would be extremely valuable as a guide to a grand unified theory beyond the SM.

\section{Cosmological defects and inflation}
\label{sec:inflation}

Metastable cosmic strings require two steps of spontaneous symmetry
breaking. In the first step, an $\text{SU(2)}$ group, which may be
embedded in some GUT group, is broken to $\text{U(1)}$,
leading to monopoles as topological defects.  In the second step, a
$\text{U(1)}$ group is spontaneously broken leading to strings
as topological defects.
This $\text{U(1)}$ group must not be orthogonal
to the $\text{U(1)}$ contained in $\text{SU(2)}$ in order to allow the
string to split into segments having monopoles and antimonopoles at
the ends.

Between the $\text{SU(2)}$ and $\text{U(1)}$ phase transitions an
inflationary period must have taken place in order to dilute the
produced monopoles but to keep the cosmic strings. In fact, one
of the motivations for cosmic inflation has been the ``monopole
problem'' of GUTs in standard cosmology (see, e.g., Ref.~\cite{Kolb:1990vq}).
One may worry that a mass ratio of $\sqrt{\kappa} \sim 8$ between the topological defects sourced by these phase transitions does not leave enough space for an inflationary period. However, while the chronological order of the symmetry breaking stages in hybrid inflation is set by the Higgs masses, these are linked to the corresponding symmetry breaking scales (or Higgs vacuum expectation values) through Yukawa couplings. This leaves enough freedom to implement hybrid inflation~\cite{Buchmuller:2021dtt}.

In the supersymmetric  $\text{SU}(2)_R\times\text{U}(1)_{B-L}$ model
discussed in Sec.~\ref{sec:metastrings}, inflation is naturally
realized by means of $F$-term hybrid inflation
\cite{Copeland:1994vg,Dvali:1994ms}. The two singlets, needed in the
superpotential to ensure $\text{SU}(2)_R$ and $\text{U}(1)_{B-L}$ breaking,
play the role of inflatons. In combination with the supersymmetric
SM and right-handed neutrinos, a consistent picture of
inflation, leptogenesis and dark matter is obtained for a large scale
of $B\!-\!L$ breaking, $v_s \sim 10^{15}~\text{GeV}$
\cite{Buchmuller:2012wn,Buchmuller:2019gfy,Afzal:2022vjx}.
Alternatively, one can consider sneutrino tribid inflation in a gauged $\text{U}(1)_{B-L}$
extension of the supersymmetric SM. Metastable strings are
again obtained by embedding the model in $\text{SO}(10)$ \cite{Masoud:2021prr}.
Depending on the pattern of supersymmetry breaking, one obtains
gravitino dark matter \cite{Afzal:2022vjx}. Note that in all $\text{SO}(10)$
models the precise connection between GUT masses and couplings and the
monopole-mass--string-tension ratio
$\sqrt{\kappa}$ is an open question that remains to be investigated.

Monopoles and strings have also been considered in nonsupersymmetric
$\text{SO}(10)$ models with an intermediate string scale
$v_s \sim 10^{13}~\text{GeV}$. The inflaton is introduced as a
GUT-singlet scalar field whose potential is generated by radiative corrections.
Monopoles and strings may be present today at an observable level, and
stochastic GWs may respect PTA and LVK bounds and only become visible
at LISA, SKA, BBO and ET/CE \cite{Chakrabortty:2020otp}. The formation of
monopole--antimonopole--string configurations may lead to a suppression
of the GW spectrum at PTA frequencies \cite{Lazarides:2022jgr, Maji:2022jzu}.

In GUT models with monopoles and metastable strings the incorporation of 
inflation is of crucial importance. One is then faced with the
challenging problem to treat gauge coupling unification, fermion
masses, proton decay, baryogenesis, (potentially) supersymmetry
breaking and dark matter, together with inflation and the formation of
a cosmic string network in a quantitatively consistent way. In the
examples mentioned above, some progress has been made, but there
is much room for improvement. Evidence for (metastable) strings
from a SGWB would be a key element to guide us toward grand unified theories.

\section{Conclusions and outlook}
\label{sec:conclusions}

The evidence for a gravitational-wave background at nanohertz frequencies recently reported by PTAs around the globe~\cite{NANOGrav:2023gor,Antoniadis:2023ott,Reardon:2023gzh,Xu:2023wog} opens a new window to study the evolution of our universe. The observed signal at nanohertz frequencies, ten orders of magnitude below the LIGO--Virgo--KAGRA band, may have an astrophysical origin\,---\,inspiralling supermassive black-hole binaries\,---\,but it might also be a remnant of events in the early universe.

One possible cosmological interpretation of the observed signal are metastable cosmic strings, which have a strong theoretical motivation in the framework of GUTs. The corresponding GW spectrum is characterized by two parameters: the string tension $\mu = 2\pi v^2_s$, where $v_s$ is the associated symmetry breaking scale, and the ratio between monopole mass squared and string tension, $\kappa = m^2_M/\mu$, which determines the lifetime of the string network.

The signal in the most recent PTA data sets~\cite{Reardon:2023gzh,NANOGrav:2023gor,Antoniadis:2023ott} is well described by a power law with a characteristic amplitude of the order of $10^{-10} \lesssim \Omega_\text{gw} h^2 \lesssim 10^{-9}$ and a positive spectral tilt, $0 \lesssim n_t \lesssim 3$, around the current PTA peak sensitivity of roughly 3~nHz. In terms of the parameters of metastable cosmic strings, this implies a string tension $10^{-11} \lesssim G\mu \lesssim 10^{-7}$ and a decay parameter $\sqrt{\kappa} \gtrsim 8$, where the upper (lower) bound on $G\mu$ ($\sqrt{\kappa}$) arises from the constraints set by ground-based interferometers on the amplitude of the SGWB. As illustrated in Fig.~\ref{fig:nt}, a value of the spectral tilt $n_t \gtrsim 1$, as preferred by the most recent PTA data, favours values $v_s \geq \text{ few } \times 10^{14}$~GeV, close to the GUT scale. Such large values of the string tension will be conclusively tested once the LIGO and Virgo ground-based interferometers reach design sensitivity in the coming years~\cite{KAGRA:2021kbb}.

In order to distinguish metastable cosmic strings from other interpretations of the SGWB signal at PTA frequencies, a more precise
determination of the spectral tilt will be important. Moreover, like most other cosmological signals, the SGWB from metastable cosmic strings is largely isotropic, as opposed to the significant anisotropies, and the possible presence of resolvable sources, which are expected for a GW signal from supermassive black-hole binaries~\cite{NANOGrav:2023tcn,Antoniadis:2023aac,NANOGrav:2023pdq}. Future pulsar observations and combinations of existing PTA data sets will shed light on these questions in the near future. In addition, GW observations in other frequency bands are an extremely powerful probe of the cosmic-string hypothesis, as the predicted signal spans many orders of magnitude in frequency: smaller frequencies would be valuable to test the characteristic $f^2$  behaviour of the spectrum, current and future ground-based detectors will be able to distinguish GUT-scale metastable strings from intermediate-scale stable strings, and the space-based interferometer LISA will probe string tensions down to values well below the current reach of PTAs. 
If upcoming observations point to an astrophysical origin of the current PTA signal, the results presented here can be interpreted as upper bounds on $G\mu$ and $\kappa$, demonstrating the potential of ground- and space-based interferometers to probe the remaining parameter space of GUT-scale cosmic strings.

On the theoretical side, the calculation of the GW spectrum has to be improved in several ways. Most importantly, a precise and robust understanding of the substructure of string loops is crucial for the estimation of the SGWB. In addition, the large value of $v_s$ suggested by the recent PTA data calls for further explicit studies of metastable strings in GUT models. As explained in Section~\ref{sec:metastrings}, the value $\sqrt{\kappa} \simeq 8.3$ hinted at by the data requires the energy scales $v_u$ and $v_s$ of the symmetry breakings leading to monopoles and strings, respectively, to be close to each other. This is a strong constraint on GUT model building that remains to be investigated, with consequences for neutrinos, leptogenesis and inflation.

\hfill

\noindent {\large \textbf{Acknowledgments}} \medskip

\noindent The work of K.\,Sc.\ is supported by the Deutsche Forschungsgemeinschaft (DFG) through the Research Training Group, GRK 2149: Strong and Weak Interactions\,---\,from Hadrons to Dark Matter.

\bibliographystyle{JHEP}
\bibliography{nanoGUT}

\end{document}